\documentclass[aps,preprint,showpacs, showkeys]{revtex4}%
\usepackage{amsfonts}
\usepackage{amsmath}
\usepackage{amssymb}
\usepackage{graphicx}
\usepackage{epsfig}
\usepackage{exscale}
\usepackage{float}
\usepackage{bbm}%
\providecommand{\U}[1]{\protect\rule{.1in}{.1in}}

\begin{document}
\title[ ]{Extrapolating the precision of the Hypergeometric Resummation to Strong
couplings with application to the $\mathcal{PT-}$Symmetric $i\phi^{3}$ Field
Theory }
\author{Abouzeid M. Shalaby}
\email{amshalab@qu.edu.qa}
\affiliation{Department of Mathematics, Statistics, and Physics, Qatar University, Al
Tarfa, Doha 2713, Qatar}

\begin{abstract}
In PRL 115, 143001 (2015), H. Mera et al. developed a new simple but precise
Hypergeometric Resummation technique. In this work, we suggest to obtain half of the
parameters of the Hypergeometric function from the strong coupling expansion
of the physical quantity. Since these parameters are taking now their exact
values they can improve the precision of the technique for the whole range of
the coupling values. The second order approximant $\, _{2}F_{1}$ of the algorithm is applied to resum the perturbation series of the ground state energy of the $\mathcal{PT-}$symmetric $(i\phi^{3})_{0+1}$ field theory. It gives accurate
results compared to exact calculations from the literature specially for very
large coupling values. The $\mathcal{PT-}$ symmetry breaking of the Yang-Lee
model has been investigated where third, fourth and fifth orders were able to
get very accurate results when compared to other resummation methods involving
$150$ orders. The critical exponent $\nu$ of the $O(4)$-symmetric model in three dimensions has been precisely obtained using only first order of perturbation series as input.  The algorithm can be extended easily to accommodate any order of perturbation series in using the generalized Hypergeometric function $\,_{k+1}F_{k}$ as it shares the same analytic properties with $\, _{2}F_{1}$.

\end{abstract}
\maketitle

In quantum field theories one is always confronted with perturbation series of
zero radius of convergence \cite{zinjustin}. For such cases Resummation
techniques have been applied and successful results have been
obtained\cite{zinjustin,zin-borel,kleinert,kleinert2}. Although these
techniques can produce accurate results they sometimes use calculations
for large number of loops as an input and most of the calculations go
numerically. Recently, a precise as well as simple Resummation technique has
been introduced \cite{Prl} where it uses only four orders of the perturbation
series as well as it is of analytic form. Such algorithm is very suitable for
quantum field cases as it can give reasonable results with only few orders out
of the perturbation series. The authors of Ref. \cite{Prl} have extended the
technique to employ higher orders of the perturbation series \cite{Prd-GF} via
use of the generalized Hypergeometric functions followed by Borel transform
that leads to a form of Meijer-G function \cite{HTF}. Regarding the original Hypergeometricc
version in Ref.\cite{Prl} and its upgrades, the prediction of the whole
parameters of the Hypergeometric function is obtained from just weak-coupling data. Since for original version of the Hypergeometric approximants, the first two parameters of the Hypergeometric function are related to its asymptotic form
for large values of the argument \cite{HTF} (Strong-Coupling data), the prediction of these
parameters from small coupling information might not lead to well known strong
coupling behaviors of the physical quantity. To employ all the information from weak-coupling as well as strong coupling expansions,  one needs to employ a generalized form of the Hypergeometric function. In this work, we introduce an algorithm that can lower the number of
input perturbation terms to half of those needed by the Hypergeometric algorithm in
Refs.\cite{Prl} as well as guarantee the accuracy for very large coupling values.

To motivate for this work, we mention that in Ref.\cite{zin-borel} the strong
coupling behaviors have been stressed for both the $\mathcal{PT-}$symmetric
and the real cubic anharmonic oscillator. In applying resummation techniques
that involve $150$ orders for the $\mathcal{PT-}$ symmetric case the authors
showed that:%
\begin{equation}
\lim_{g\rightarrow\infty}\frac{E_{0}}{\left\vert g\right\vert ^{\frac{1}{5}}%
}=0.372545790452207098250601(1), \label{lpt}%
\end{equation}
while for the cubic oscillator they obtained
\begin{align}
\lim_{g\rightarrow-\infty}\frac{E_{0}}{\left\vert g\right\vert ^{\frac{1}{5}%
}}  &  =0.3013958756586835717823(7)\nonumber\\
&  +0.2189769214314493762936(0)i \label{Lco}%
\end{align}
It is easy to check that the prediction of the original Hypergeometric
Resummation in \cite{Prl} gives zero in both cases. This is because there is
like $6\%$ error in the prediction of the second parameter in the
Hypergeometric function which affects the precision of the algorithm for large
coupling values. Accordingly, one needs to extrapolate the predictions of the
algorithm to give accurate results for the whole coupling space.

In this work we apply the Hypergeometric Resummation algorithm to the
$\mathcal{PT-}$ symmetric Yang-Lee model but in guiding the Hypergeometric
functions with parameters from the strong coupling behavior. In $0+1$
space-time dimension (quantum mechanics), one can follow a scaling as well as
gauge canonical transformations to obtain the strong coupling expansion of the
theory \cite{zin-borel}. In higher dimensions (quantum field theory), for some cases  one can obtain
 the strong-coupling expansions of a physical quantity \cite{bend-strong, keinert-strong}. So feeding the resummation technique with
two parameters (for the second order) from the perturbation series and the
other two from the strong coupling expansion is possible for both quantum and
quantum field problems. As we will see in this work, this algorithm lowers the
number of orders from the perturbation series to two instead of four needed
for the original algorithm in Ref.\cite{Prl}. Besides the prediction is then
more accurate for large couplings. The extension of the algorithm to higher
orders is direct and shall be applied here to investigate $\mathcal{PT-}$
symmetry breaking of the Yang-Lee model with increasing precision when moved
from $_{2}F_{1}$ to $_{3}F_{2}$ and very high precision obtained from
$_{4}F_{3}$ and $_{5}F_{4}$ when compared to resummation results from
Ref.\cite{zin-borel} where methods there involved $150$ orders. It is worth
mentioning that the strong coupling parameters have been used before in the
literature to accelerate the convergence of resummation algorithms
\cite{Strong1,Strong2,Strong3,Strong4,Strong5,Strong6,Strong7} and thus it is
expected to have the same role in the Hypergeometric resummation algorithm too.

The Yang-Lee model or equivalently $\mathcal{PT-}$symmetric \ $i\phi^{3}$
field theory has been exposed to recent discussions because it has an
imaginary potential but on the other hand has a real spectrum \cite{Bend-ptsv,
bend-lo, Benderx3, Zadahx3,Shin1,Shin2}. In fact, the ground state energy has a
zero radius of convergence and thus non-perturpative Resummation algorithms
are in a need to get reliable results from perturpative calculations as an
input. Another aspect for which non-perturbative approaches of the Yang-Lee
model are essential is because it represents the Landau-Ginzberg approximation
of the Ising model near the Yang-Lee edge singularity \cite{lee1,lee2,LYsing1,Musardob,Dorey2}. Pad$\acute{e}$, Borel and other
algorithms applied to the model in Refs.\cite{pade,spect,crossing,prdphi3} .
While Pad$\acute{e}$ approximation can not account for the strong coupling behavior,
most of Borel calculations are achieved via numerical calculations. On the other hand, the recent
Hypergeometric Resummation technique introduced in \cite{Prl} is characterized
by being simple, closed form as well as employs only few number of terms from
the perturbation series as an input. In $0+1$ space-time dimensions (quantum
mechanics) the Yang-Lee model has been stressed by the authors of Ref.
\cite{Prl} but we realized that the precision of the results for strong
coupling (even for the real potential case) is questionable. As we suggested
above, a way to have better fitting with available exact results is to provide
the Hypergeometric function with known results from the strong coupling
behavior. In fact, strong coupling expansion can be obtained in many cases.
For instance, Hermitian theories like the $\phi^{4}$ field theory has been
extensively stressed in the literature and its strong coupling as well as
large order behaviors are discussed and employed to accelerate the convergence
of resummation techniques\cite{zinjustin,keinert-strong}. Although for quantum
field theory (dimension $d>1$) the strong coupling parameters can be obtained
using optimization methods, the employment of their approximate values lead to
improvement of the resummation results \cite{
keinert-strong,keinert-strong2,Strong0}.

Applying a simple and accurate Resummation algorithms to  divergent sreis  might have
a strong impact on the field of $\mathcal{PT-}$ symmetric field theories where
one can resum the series from the known results of just the first few terms in
the perturbation series and its strong coupling behavior. For instance, the $\mathcal{PT-}%
$symmetric ~$(-\phi)^{4}$ model is assumed to be asymptotically free
\cite{aboebt, Symanzik,bendf,Frieder} but up to the best of our knowledge no
non-perturbative calculation for the Beta function appeared yet. Another
application that the Hypergeometric Resummation can play a vital role is in
the very recently introduced $\mathcal{PT-}$symmetric Higgs Mechanism and such
Resummation technique may offer a non-perturbative tool that saves the effort
and time for the calculation in such cases where high order of loop
calculations is time consuming. A note to be mentioned is that it took the researchers like  $25$ years to move from the fifth order to sixth order of the perturbation sires of the renormalization group functions for simple quantum field theories \cite{25I,25II}. So seeking a way to accelerate the convergence of a resummation algorithm is more than important in the field of quantum field theory.  As we will see in this work, the critical exponent of the $O(4)$-symmetric field theory is obtained at only the first order using the algorithm in this work. This theory can describe the finite temperature phase transition in QCD   with two light flavors \cite{QCD-O4}.

For the Hypergeometric Resummation there is another precision realization
concerning the small coupling predictions where it has been realized by the
authors themselves in Ref. \cite{cut}. According to them, the series expansion
of the Hypergeometric function does not have a zero radius of convergence
while the aim is to sum a series of zero radius of convergence. To solve this
problem, the authors set an algorithm that results in a Hypergeometric
Resummation with zero radius of convergence \cite{cut}. In this work, we shall
stress only the impact of employing the strong coupling behavior on the
accuracy of the algorithm.

The Hypergeometric function $_{2}F_{1}$ can have a power law behavior near
singular points \cite{HTF} and thus in principle can account for the
calculation of the critical exponents of the Yang-Lee model near the edge
singularity but this will be out of the scope of this work.

To test the accuracy of algorithm before we go to the $\mathcal{PT-}$symmetric
$i\phi^{3}$ theory, we consider  the $\,_{2}F_{1}$ resummation of the
ground state energy of the anharmonic oscillator where it has a perturbation
series with zero radius of convergence \cite{bend-anhar}. We obtained the
ground state energy from second order approximant as $E_{0}\left(  g\right)  =\,_{2}F_{1}\left(  \frac{1}%
{3},\frac{-1}{3},c,-dg\right)  $ and find $E_{0}\left(  50\right)
=2.4484029106721046$. Although this result is better than Borel-Pad$\acute{e}$
resummation in using $24^{th}$ order $BP12/12$ where it gives $E_{0}\left(
50\right)  =2.3157388197$ \cite{Prd-GF}, one can get better results in
involving more terms from the perturbation series. When we use $\,_{4}F_{3}$
we get $E_{0}\left(  50\right)  =2.4856072532925255$. Note that, for this
theory, scaling properties can lead to the strong coupling expansion for the
ground state energy of the form:%

\begin{equation}
E_{0}=g^{\frac{1}{3}}\sum_{i=0}^{\infty}m^{2}\left(  g^{\frac{-2}{3}}\right)
^{i}.
\end{equation}
Accordingly, the $a_i$ parameters in the generalized Hypergeometric approximant $_{p}F_{p-1}(a_{1,}a_{2},....a_{p};b_{1}%
,b_{2,}.......b_{p-1};-\sigma z)$ are given by $a_{1}=\frac{-1}{3},a_{2}=\frac{1}{3},a_{3}=1,a_{4}=\frac{5}%
{3},a_{5}=\frac{7}{3},..........$ Note also that the exact result for
$E_{0}\left(  50\right)  $ is $2.4997087726$ and the Hypergeometric
resummation in Ref.\cite{Prd-GF} gives $2.4997107287$ but in involving $25$
orders. So it seems that the algorithm we use accelerates the convergence to
the exact prediction as the second order in our algorithm have accuracy that
lies between the $24^{th}$ order of Borel-Bad$\acute{e}$ resummation and the $25^{th}$
of the generalized Hypergeometric (Meijer-G)  resummation in Ref.\cite{Prd-GF} while our
fourth order result is very close to the exact one.  

\begin{figure}[t]
\begin{center}
\epsfig{file=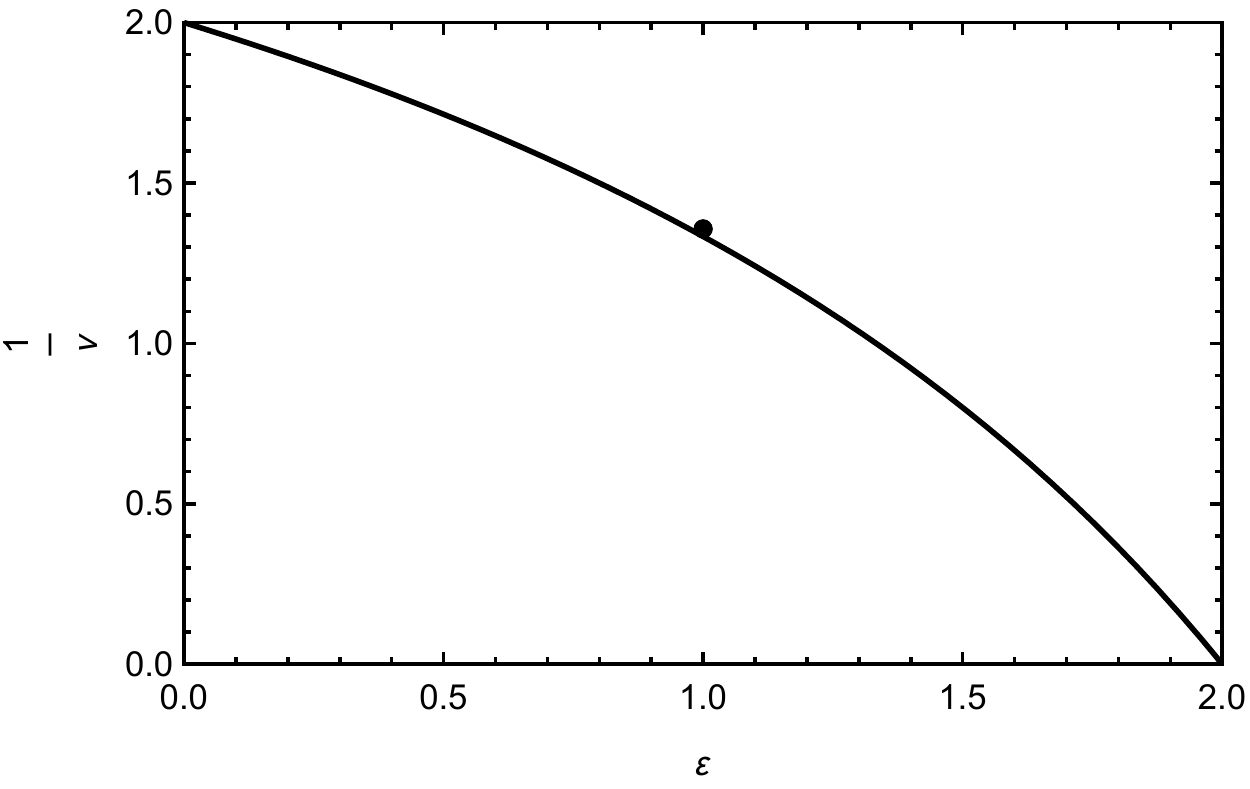,width=0.65\textwidth}
\end{center}
\caption{\textit{The approximant $\nu^{-1}= 2 \   _{1}F_{0}(1, \, \frac{-1}{4}  \tilde{\varepsilon})$ versus $\varepsilon=4-D$ for the $O(4)$-symmetric model with dot represents the six loops prediction for comparison with Figure 3 in Ref.\cite{Klein-sig} }.}%
\label{nu}%
\end{figure}

Another example to test   algorithm is to stress the critical exponent $\nu$ of the $O(4)$-symmetric model. In Ref.\cite{Klein-sig}, the $\varepsilon$-  expansion of the critical exponent $\nu$ of the $O(4)$ model ($\varepsilon$=4-D) has been linked to  the $\epsilon$-expansion of the $\sigma$-model ($\epsilon=2-D)$ using the fact that both lie in the same class of universality. In fact, the authors were able to write one expansion as the strong coupling expansion of the other in writing the two expansions in terms of a new variable $\tilde{\varepsilon}=2\frac{D-4}{D-2}$. They  obtained the strong coupling   expansion (N=4) as:
\begin{equation}
\nu ^{-1}=4\tilde{\varepsilon}^{-1}-8\frac{N-4}{N-2}\tilde{\varepsilon}%
^{-2}+O(\tilde{\varepsilon}^{-3})
\end{equation}
while the weak coupling expansion is given by:
\begin{equation}
\nu ^{-1}=2-\frac{1}{2}\tilde{\varepsilon}+0.0833333\tilde{\varepsilon}%
^{2}+O(\tilde{\varepsilon}^{3}).
\end{equation}
In fact, the $a_i$ parameters in the approximants  $_{p}F_{p-1}(a_{1,}a_{2},....a_{p};b_{1}%
,b_{2,}.......b_{p-1};\sigma z)$  can be concluded from the strong coupling expansion to be $1,2,3,.....$ and so on. However, one can realize that in that case $a_i-a_j$ is an integer which means that the strong coupling expansion of $_{p}F_{p-1}(a_{1,}a_{2},....a_{p};b_{1}%
,b_{2,}.......b_{p-1};\sigma z)$ will include logarithmic dependance \cite{ HTF} which is not reflected in the strong coupling expansion above. The only exception is for the approximant $_{1}F_{0}(1,\ , \  \sigma  \tilde{\varepsilon})$ . Matching the expansion of this approximant  with the weak-coupling expansion above one finds that $\sigma=-\frac{1}{4}$. Accordingly, the resummed $\nu$ exponent is given by 
	\[
	\nu=\frac{1}{2 \   _{1}F_{0}(1, / \frac{-1}{4}  \tilde{\varepsilon})}.
\]
In three dimensions ($\varepsilon=1$)  it gives the result $\nu=0.75$ compared to $0.735$ from the resummation of fifth order series in Ref.\cite{Klein-sig} and  $0.7479$ using MonteCarlo sc.  \cite{nu-montecarlo,Pelissetto}.  Note that we get this result using only first order from perturbation series as input which up to the best of our knowledge is the first time to get that result that fast. To compare with figure 3 in Ref. \cite{Klein-sig}, we generated the graph in Fig.\ref{nu} while the details of the critical exponents of $O(N)$-symmetric model using the Hypergeometric algorithm is postponed to a separate work. 

Ironed by the precise results we obtained above from a very few orders of perturbation series as input, we  stress a model that is can be considered as a hot research point \cite{zin-borel,Zadahx3,Shin1,Shin2,abox3,x315,x3-4L}.  That model is  the $\mathcal{PT-}$symmetric $i\phi^{3}$ theory which has a Lagrangian density of the form: 
\begin{equation}
\mathcal{L}\left[  \phi\right]  =\frac{1}{2}\left(  \partial\phi\right) 
^{2}-\frac{1}{2}m^{2}\phi^{2}(x)-\frac{i\sqrt{g}}{6}\phi^{3}\left(  x\right)
, \label{lag}%
\end{equation}
with a corresponding Hamiltonian density:%
\begin{equation}
H=\frac{1}{2}\pi^{2}+\frac{1}{2}\left(  \nabla\phi\right)  ^{2}+\frac{1}%
{2}m^{2}\phi^{2}(x)+\frac{i\sqrt{g}}{6}\phi^{3}\left(  x\right)  .
\label{iphi3}%
\end{equation}
The Hamiltonian operator is $\mathcal{PT-}$symmetric and thus the spectrum is
real. This Hamiltonian is closely related to the Hamiltonian
\begin{equation}
H_{J}=\frac{1}{2}\pi^{2}+\frac{1}{2}\left(  \nabla\phi\right)  ^{2}+\frac
{i}{6}\phi^{3}\left(  x\right)  +iJ\phi, \label{Yang}%
\end{equation}
where in $0+1$ space-time dimensions, one can start from $H$ and apply scaling
as well as gauge transformation to get $H_{J}$ \cite{zin-borel,
spect,x3-equiv}. Although $H_{J}$ is $\mathcal{PT-}$symmetric, the
$\mathcal{PT-}$symmetry is not exact for all real $J$ values and the
$\mathcal{PT-}$symmetry is broken for some critical $J$ value \cite{zin-borel}%
. According to Yang and Lee, the partition function or equivalently the vacuum
to vacuum amplitude can have a zero for negative $\mathcal{J}$ values known as
Yang-Lee edge singularity \cite{lee1,lee2,LYsing1,Musardob}. This specific
zero is associated with non-analyticity of the ground state energy and this is
supposed to be associated with$\mathcal{PT-}$symmetry breaking. Near the edge
singularity the theory is totally non-perturbative and one needs to apply
non-perturbative techniques. The perturbation series for the ground state
energy of the Hamiltonian $H$ also has a zero radius of convergence and thus
Resummation is needed anyway. We will stress the resummation of ground state
energy of both Hamiltonians $H$ and $H_{J}$ and show that our second to fourth
orders are very competitive to the results with the $150^{th}$ order of
resummation methods in Ref.\cite{zin-borel}. For $H_{J}$, the important
$\mathcal{PT-}$ symmetry breaking near the edge singularity will be shown in
our results while the first $20^{th}$ orders of weak coupling expansion cannot
account for it.

For the model represented by the above Lagrangian density, up to first order
in $g$, the ground state energy receives only contributions from the sunset
and dumbbell diagrams shown in Fig.\ref{sunset} while Mercedes and other four
diagrams contribute to the $g^{2}$ order which are shown in Fig.\ref{g2}
(these diagrams are all listed in Ref.\cite{Bend-ptsv} too). Accordingly, up
to $g^{2}$ order, the vacuum energy is given by:

\begin{figure}[htbp]
\begin{center}
\epsfig{file=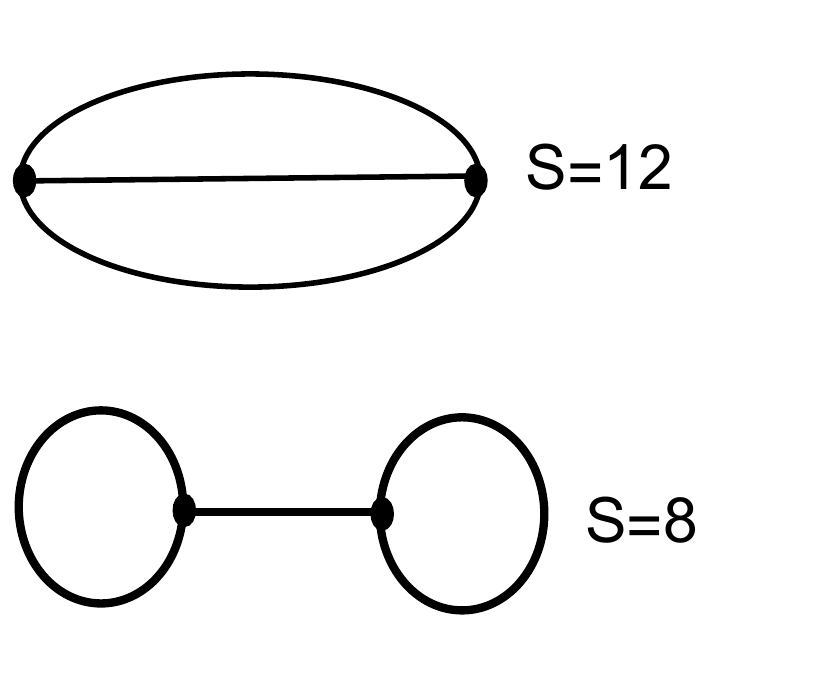,width=0.65\textwidth}
\end{center}
\caption{\textit{The 2-vertices Feynman diagrams contributing to the first
order in squared-coupling $g$ of vacuum energy for the $\mathcal{PT-}%
$symmetric $(i \phi^{3})_{0+1}$ field theory. The symmetry factor $S$ is
written for each diagram.}}%
\label{sunset}%
\end{figure}

\begin{figure}[t]
\begin{center}
\epsfig{file=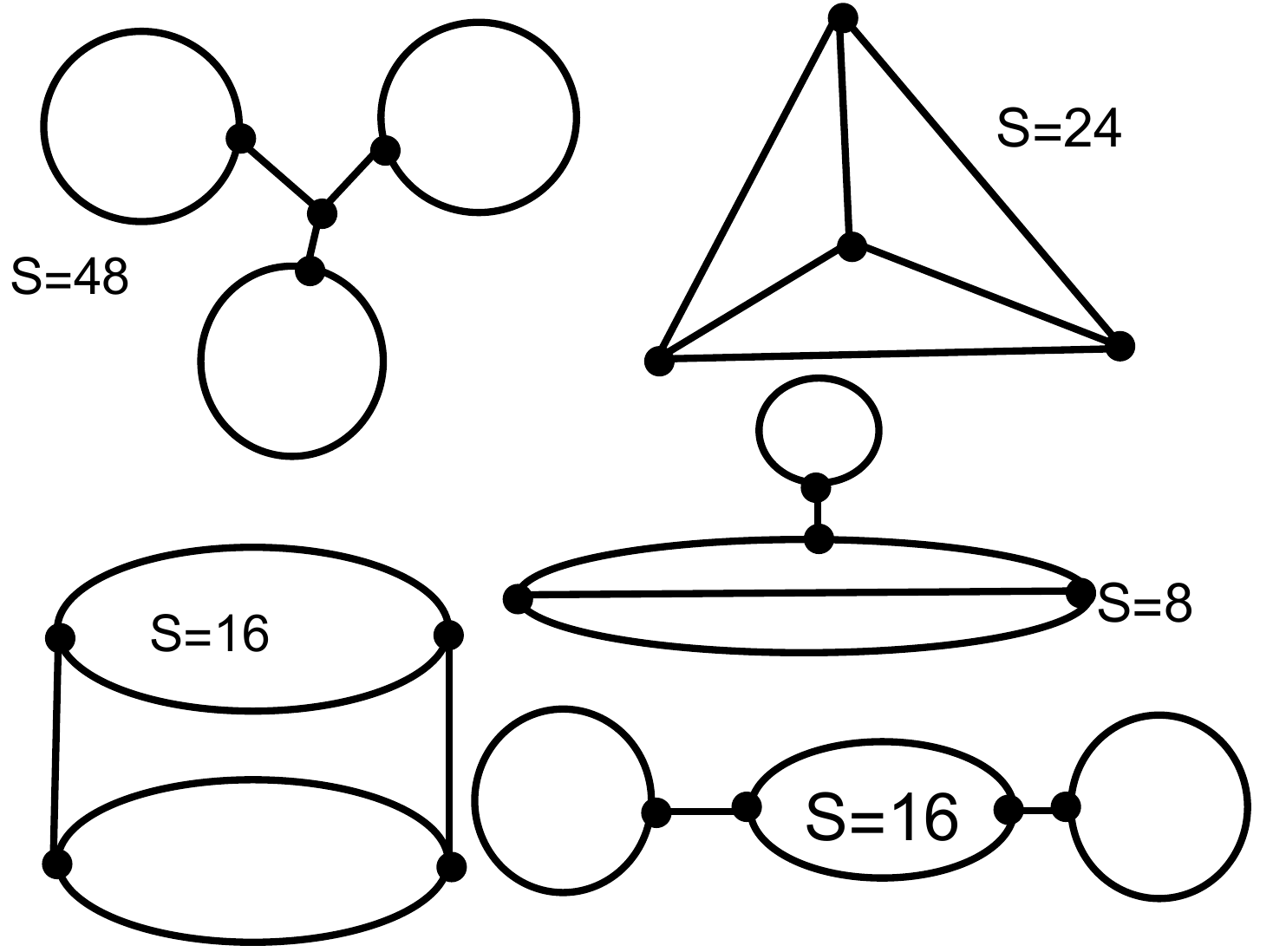,width=0.65\textwidth}
\end{center}
\caption{The vacuum diagrams contribution to the order $g^{2}$ of the ground
state energy of the $\mathcal{PT-}$symmetric $i\phi^{3}$ field theory.}%
\label{g2}%
\end{figure}%
\begin{equation}
E_{0}=\frac{1}{2}+\frac{11g}{288}-\frac{930}{288^{2}}g^{2}, \label{pertub}%
\end{equation}
where we assumed in Eq.(\ref{lag}) that $m=1$ and the space-time dimension is
$0+1$. This result can be checked in many articles although some of them
obtained it in different basis \cite{zin-borel, Bend-ptsv}. It is well known
that this series is Borel summable as well as having a zero radius of
convergence. Accordingly, non-perturbative techniques are needed in order to
get reasonable results for the vacuum energy. In many articles, different
techniques applied ranging from Pad$\acute{e}$ approximation \cite{bend-lo} and Borel
Resummation \cite{zin-borel} to the recent Hypergeometric Resummation
\cite{Prl}. In fact, Pad$\acute{e}$ approximation although can account for needed branch
cuts of the divergent series, it fails to reproduce the strong coupling
behavior of the physical quantity under consideration. 

\begin{figure}[t]
\begin{center}
\epsfig{file=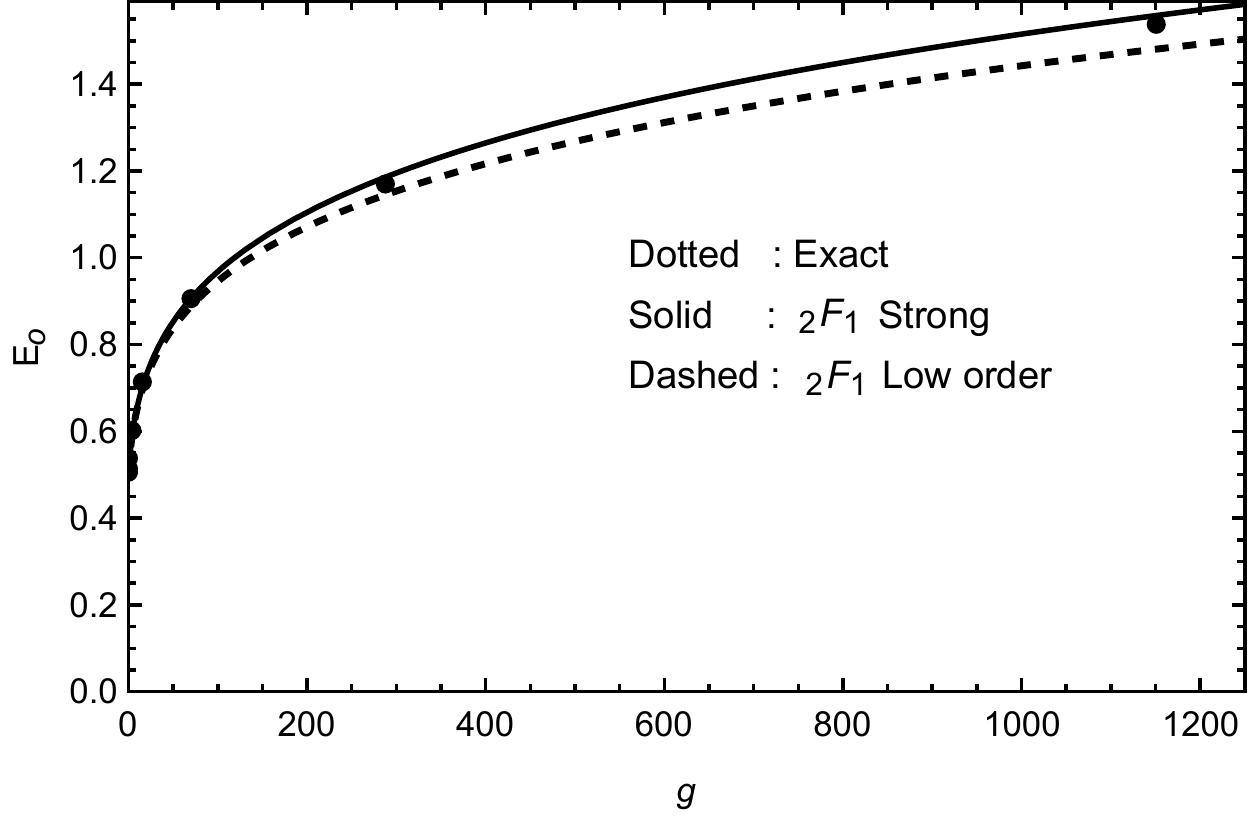,width=0.65\textwidth}
\end{center}
\caption{\textit{Comparison between our Resummation formula $_{2}F_{1}$ for
the ground state energy of the Hamiltonian in Eq.(\ref{iphi3})(solid), the
original form in Ref.\cite{Prl} (dashed) and exact results from
Ref.\cite{bend-lo} (dots). Note that the coupling in our work is rescaled from
that in Ref.\cite{bend-lo} where $g$ in our work is equivalent to
$288\lambda^{2}$ in that reference.}}%
\label{Hyper-3}%
\end{figure}

The Hypergeometric  algorithm in Ref.\cite{Prl} has features that recommend it for resummation of divergent series. The suggested  $_{2}F_{1}(a_{1,}a_{2};b_{1} ;\sigma z)$ approximant needs information from four orders in perturbation series. According to the Hypergeometric Resummation of a
divergent series, the vacuum energy of the Hamiltonian $H$ in $0+1$ space-time
dimensions is given by:%

\begin{equation}
E_{0}=\frac{1}{2}\text{ }_{2}F_{1}\left(  a,b,c,-\frac{g}{d}\right)  ,
\label{hyper-gen}%
\end{equation}
where $_{2}F_{1}$ is the Hypergeometric function and the parameters $a,b,c$
and $d$ can be obtained from the series expansion of the Hypergeometric
function and then matching them with the first four terms in the perturbation
series. The series expansion of Eq.(\ref{hyper-gen}) can be obtained as
\begin{align*}
E_{0}  &  =\frac{1}{2}-\frac{(ab)}{2(cd)}g+\frac{a(a+1)b(b+1)}{4c(c+1)d^{2}%
}g^{2}-\frac{(a(a+1)(a+2)b(b+1)(b+2))}{12\left(  c(c+1)(c+2)d^{3}\right)
}g^{3}\\
&  +\frac{a(a+1)(a+2)(a+3)b(b+1)(b+2)(b+3)}{48c(c+1)(c+2)(c+3)d^{4}}%
g^{4}+O\left(  g^{5}\right)
\end{align*}
The order of the series in Eq.(\ref{pertub}) does not have enough information
to solve for the four unknown parameters. So one have to add contributions
from vacuum diagrams up to eight vertices. Although this is possible but in
applying the method to a more realistic field theory like QCD, it will be time
consuming. It would be better to seek a way to lower the number of orders in
the perturbation series needed to find the different parameters. Rather than
this, the parameters when all are obtained from just the first few orders in
perturbation series, the resumed function does not reproduce well known strong
coupling limits of the Yang-Lee model shown in Eqs(\ref{lpt}) \& (\ref{Lco}).
So it is very necessary to feed the Hypergeometric function with parameters
obtained from strong coupling behavior. In fact, when $a-b$ is not an integer
and for large values of $\left\vert g\right\vert ,$ the Hypergeometric
function has the following asymptotic form\cite{HTF};%
\[
_{2}F_{1}\left(  a,b,c,g\right)  \sim\lambda_{1}g^{-a}+\lambda_{1}%
g^{-b},\left\vert g\right\vert \gg1.
\]
The strong coupling behavior of a physical quantity can be obtained exactly in
some cases. In such cases, the parameters $a$ and $b$ are known exactly and
only a second order of the perturbation series is sufficient to predict the
other two parameters. In $0+1$, the Hamiltonian takes the form
\begin{equation}
H=\frac{1}{2}\pi^{2}+\frac{1}{2}m^{2}\phi^{2}(x)+\frac{i\sqrt{g}}{6}\phi
^{3}\left(  x\right)  .
\end{equation}
A symmetry transformation of the form
\begin{align*}
\phi  &  \rightarrow\exp\left(  -w \pi\right)  \phi\exp\left(  w\pi\right)  =\phi-w\left[
\pi,\phi\right] \\
&  =\phi+iw,\text{ }w=\frac{m^{2}}{\sqrt{g}},
\end{align*}
leads to:
\begin{equation}
H=\frac{1}{2}\pi^{2}+\frac{1}{2}\left(  \nabla\phi\right)  ^{2}+\left(
\frac{1}{6}i\sqrt{g}\right)  \phi^{3}+\frac{im^{4}}{2\sqrt{g}}\allowbreak
\phi-\frac{m^{6}}{3g}\
\end{equation}
Note that this transformation changes the metric operator but keeping the
spectrum invariant \cite{canonical,abo-iso}. If we follow this by a scaling
transformation of the form $\exp\left(  i\ln\beta\right)  $ it scales $\phi$ by
a factor $\beta$ \cite{canonical}. Taking $\beta=g^{-\frac{1}{10}}$ leads to
the result:%
\begin{equation}
H=\sqrt[5]{g}\left(  \frac{\pi^{2}}{2}+\frac{i\phi^{3}}{6}+\frac{1}{2}%
\frac{im^{4}}{g^{\frac{4}{5}}}\phi\right)  -\frac{m^{6}}{3g}.
\end{equation}
This form suggests an expansion for the energy in the form
\begin{equation}
E_{0}=-\frac{m^{6}}{3g}+\ \ \sum_{l=0}^{\infty}c_{l}g^{-\frac{4l-1}{5}}.
\label{LOP}%
\end{equation}
This result has been shown in Ref.\cite{zin-borel} and it suggests that $a=1$
while $b$ equals $\frac{-1}{5}$. Accordingly, the Hypergeometric Resummation
of the perturbation series in Eq.(\ref{pertub}) takes the form
\begin{equation}
E_{0}=\frac{1}{2}\text{ }_{2}F_{1}\left(  1,\frac{-1}{5},c,-\frac{g}%
{d}\right)  . \label{Resumm}%
\end{equation}
The parameters $c$ and $d$ can be found from matching the coefficients of the
first two terms from series expansion of this equation with those in
Eq.(\ref{pertub}) and then we get:
\[
E_{0}=\frac{1}{2}\,_{2}F_{1}\left(  1,-\frac{1}{5},\frac{465}{19},-\frac
{8525}{912}g\right)  .
\]
The resummed form in Eq.(\ref{Resumm}) is fed with information from small
coupling (parameters $c\&d$) and strong coupling (parameters $a\&b$)
behaviors. Accordingly, one expect to give accurate results for the whole
range of the coupling space. To test that expectation, let us check the
accuracy of the Resummation formula in Eq.(\ref{Resumm}). For $g=\frac{1}{2}$,
we get $E_{0}=0.516915$ compared to the best of the resummation algorithms at
$150^{th}$ order in Ref.\cite{zin-borel} which gives $E_{0}=0.516892$. Also, for
$g=1$, we have $E_{0}=0.530886$ compared to $E_{0}=0.5307818$ form
Ref.\cite{zin-borel}. Now we compare with some larger values of $g$. For
$g=\frac{288}{49}$, one gets $E_{0}=0.614319$ while the result in
Ref.\cite{zin-borel} gives $E_{0}=0.612738.$ Also, for $g=4\times288$, we get
$E_{0}=1.55851$ while the exact value (reported in the last row in table $III$
in Ref.(\cite{bend-lo})) is $E_{0}=1.53078$. So it seems that the simple
method of Hypergeometric Resummation gives precise results though it has been
fed with information of the first two terms in the perturbation series.
However, the original version introduced in Ref.\cite{Prl} gives precise
results for a wide range of coupling values but not for very large coupling
values. For instance when $g=4\times288$ it gives $E_{0}=1.48104$ which is not
as accurate as our prediction when both are compared with the exact value
above. For more tests of our results and also the original version introduced
in Ref.\cite{Prl} one needs to check for the limit at $g\rightarrow\pm\infty.$
For $g\rightarrow-\infty$ our prediction is
\[
\ \lim_{g\rightarrow-\infty}\frac{E_{0}}{\left\vert g\right\vert ^{\frac{1}%
{5}}}=0.30738+0.223325i
\]
while the prediction of the original form in Ref.(\cite{Prl}) is zero and the
methods in Ref.(\cite{zin-borel}) gives $0.30139588+0.2189769214i$. This is
expected because any tiny difference in the parameters $a$ and $b$ will ruin
the strong coupling behavior of the resummed function. In fact, the imaginary
part of the vacuum energy for a real potential can be obtained by
non-perturbative techniques only and thus it is always a good test for any
Resummation tool. On the other hand, for the $\mathcal{PT}$-symmetric case
$\left(  g\rightarrow\infty\right)  $, we can find the result:%

\[
\lim_{g\rightarrow\infty}\frac{E_{0}}{\left\vert g\right\vert ^{\frac{1}{5}}%
}=0.379943,
\]
while the result from Ref.\cite{zin-borel} is
\[
\lim_{g\rightarrow\infty}\frac{E_{0}}{\left\vert g\right\vert ^{\frac{1}{5}}%
}=0.3723,
\]
but as expected the original Hypergeometric Resummation algorithm gives zero
again. These results show clearly that feeding the Hypergeometric Resummation
with parameters from the strong coupling behavior is necessary to extrapolate
the prediction to the large coupling behavior of the resummed function. The
accuracy of our results for large coupling values over the predictions from
the original algorithm in Ref.\cite{Prl} is clear from Fig.\ref{Hyper-3}. In
this figure one can realize that both of our formula and original one give
reasonable results compared to exact results for not so large values of the
coupling. For very large values however, one can realize that our formula fits
well with exact results but the original formula deviates from the exact
results. This is expected as the parameters in Ref.\cite{Prl} are all
predicted from the first four terms in the perturbation series and thus
expected to loose memory for strong coupling predictions. 

\begin{figure}[pth]
\begin{center}
\epsfig{file=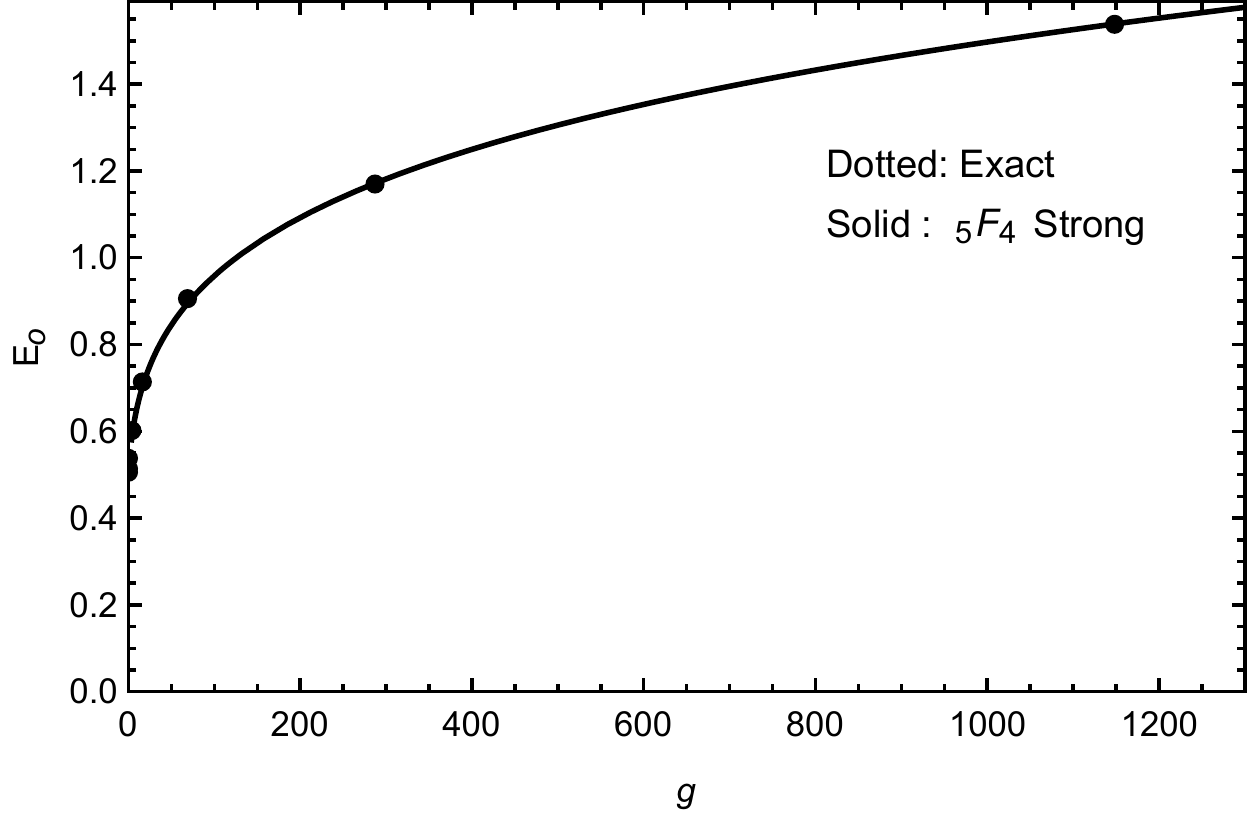,width=0.65\textwidth}
\end{center}
\caption{\textit{Comparison between our Resummation formula $\,_{5}F_{4}$ for
the ground state energy of the Hamiltonian in Eq.(\ref{iphi3})(solid) and
exact results from Ref.\cite{bend-lo} (dots). The accuracy looks improved when
compared with the second order ($\,_{2}F_{1}$) in Fig. \ref{Hyper-3}.}}%
\label{5F4g}%
\end{figure}

The extension of the method to higher orders is direct as one suggests the
resummation function to be $_{p}F_{q}(a_{1,}a_{2},....a_{p};b_{1}%
,b_{2,}.......b_{q};-\sigma z)$. When $p=q+1$, the set of functions $_{p}%
F_{q}(a_{1,}a_{2},....a_{p};b_{1},b_{2,}.......b_{q};-\sigma z)$ are all
sharing the same analytic properties. In our algorithm, the $a_{i}$ parameters
are determined exactly from the strong coupling expansion of the theory under
consideration while $b_{i}$ and $\sigma$ parameters are determined from $q+1$
set of algebraic equations obtained by comparing the series expansion of
$_{p}F_{q}(a_{1,}a_{2},....a_{p};b_{1},b_{2,}.......b_{q};-\sigma z)$ with the
perturbation series of the physical quantity. In fact, this algorithm reduces
the non linearity of the parameters equations to half. One can even get an
equivalent set of equations which are all linear in $\sigma$ and all the
equations consider only powers of one in each parameters. This strategy avoids
troubles faced in the generalized Hypergeometric resummation technique in the
literature in solving the set of equations of the $N$ parameters. For the
Hamiltonian $H$ in Eq.(\ref{iphi3}) we resummed the ground state energy using
$\,_{2}F_{1},$ $\,_{3}F_{2},$ and $\,_{4}F_{3}$ and the results are listed in
table \ref{tab:wgresum} compared to exact results and the $150^{th}$
resummation techniques from Ref.\cite{zin-borel}. It is clear that our fourth
order resummation ($\, _{4}F_{3}$) gives accurate results and the accuracy is
improved systemically when moving to higher orders. The plot of the fifth
order ($\,_{5}F_{4})$ versus exact results for the same quantity is shown is
Fig.\ref{5F4g} where one can realize that the precision of the algorithm is
improving systematically from second ($\, _{2}F_{1}$) to fifth order ($\,
_{5}F_{4}$).

\begin{table}[pth]
\caption{{\protect\scriptsize {The Hypergeometric resummation $_{2}F_{1}$,
$_{3}F_{2}$ and $_{4}F_{3}$ for the ground state function in Eq.(\ref{iphi3})
compared to the $150^{th}$ order of resummation methods in
Ref.\cite{zin-borel} and exact results. It is very clear that the second order
$_{2}F_{1}$ gives accurate results and we get higher precision in going to
higher orders where our $4^{th}$ order resummation $_{4}F_{3}$ gives results
competitive to the the $150^{th}$ order of resummation methods in
Ref.\cite{zin-borel}.}}}%
\label{tab:wgresum}
\begin{tabular}
[c]{|l|l|l|l|l|l|}\hline
$g$ & $_{2}F_{1}$ & $_{3}F_{2}$ & $_{4}F_{3}$ & $E^{150}_{0}$ \cite{zin-borel}& $E_{exact}%
$\\\hline
0.5 & 0.516915482 & 0.51689308 & 0.516891566 & 0.516891764 & ---\\\hline
1 & 0.530885535 & 0.53079024 & 0.53077974 & 0.5307817593 & 0.53078176\\\hline
$\frac{288}{49}$ & 0.614318594 & 0.61296986 & 0.61260464 & 0.61273810639 &
0.612738106\\\hline
\end{tabular}
\end{table}

 It is well known that the  series $_{p}F_{p-1}(a_{1,}a_{2},....a_{p};b_{1}%
,b_{2,}.......b_{p-1};\sigma z)$ has a finite radius of convergence while it has been used to resum a divergent series of zero radius of convergence and this issue has been stressed in Ref.\cite{cut}. However, the parameter $\sigma$ is taking large values that accounts for very small radius of convergence (but non-zero) and thus the resummation gives good results specially for not so small couplings. On the other hand, we have in physics divergent series that do have finite radius of convergence for which the Hypergeometric resummation is more suitable and expected to give more precise results. Examples of these kind are the strong coupling expansion of a physical quantity. The vacuum energy of the Yang-Lee model is of that type.    

\begin{figure}[t]
\begin{center}
\epsfig{file=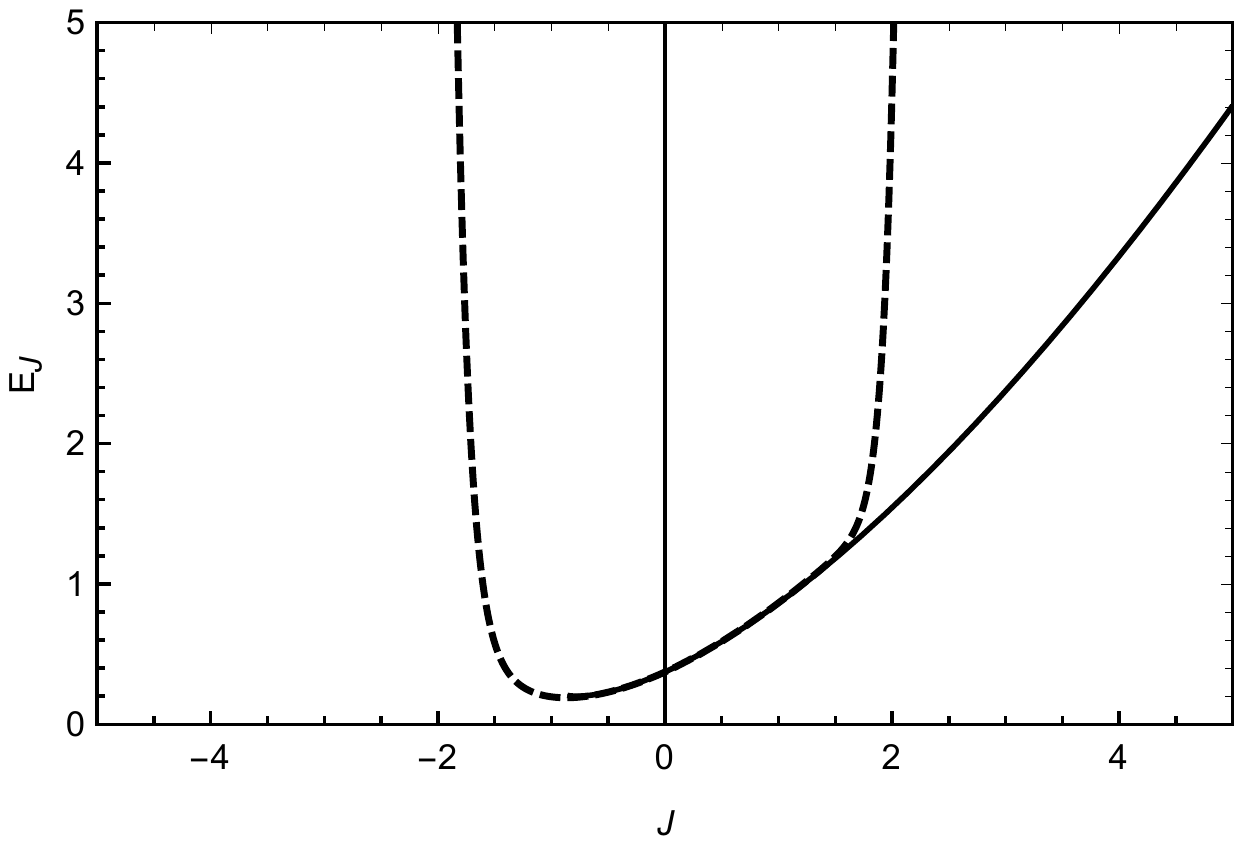,width=0.65\textwidth}
\end{center}
\caption{\textit{Comparison between our Resummation formula $_{2}F_{1}$ for
$E_{J}$ (solid) and the $20^{th} $ order of the perturbation series (dashed)
from Ref.\cite{zin-borel}. While the agreement is good for a range of the
coupling $J$, the perturbation series fails (as expected) to produce the
$\mathcal{PT}$-symmetry breaking expected as well as fails to fit with strong
coupling behavior.} }%
\label{xi}%
\end{figure}

For the investigation of $\mathcal{PT}$-symmetry breaking in the Yang-Lee model represented by the Hamiltonian $H$ in Eq(\ref{Yang}), the ground state
energy up to second order in $J$ is given by \cite{zin-borel};%
\begin{align*}
E_{J}  &  =.3725457904522070982506011+0.3675358055441936035304J\\
&  +0.1437877004150665158339J^{2}+O\left(  J^{3}\right)
\end{align*}
The resummation for this perturbation series is then obtained as:%
\[
E_{J}=\, _{2}F_{1} (-3/2,-1/4,b_{1};-\sigma J).
\]
We compared our results with the $20^{th}$ order of the perturbation series
from Ref.\cite{zin-borel} in Fig.\ref{xi}. The resumed series gives reasonable
agreement with this relatively high order of perturbation series but deviate
from each other near the critical region where the resumed formula starts to
be complex ($\mathcal{PT}$-symmetry breaking) and also separated for a
relatively strong coupling. A note to be mentioned that the resumed formula
using four terms from the perturbation series as in Ref.\cite{Prl}, agrees
well with our result but they deviates at very strong couplings as expected.
To get more accurate results to be compared with $150^{th}$ order of
resummation methods in Ref.\cite{zin-borel}, we obtained also the resummation
aproximants    $_{3}F_{2}$, $_{4}F_{3}$ and $_{5}F_{4}$ where we
listed them in table \ref{tab:wxiresum}. It is very clear that our resummation
formula are giving precise results although we used only low orders of
calculations compared to the the $150^{th}$ order of resummation methods in
Ref.\cite{zin-borel}. Note that in this table for $J=-1$, $\, _{2}F_{1} $
results in a tiny imaginary part to the ground state energy and this is
because it predicts a smaller critical coupling than $\, _{3}F_{2}$,$\,_{4}F_{3}$ and $\, _{5}F_{4}$ and the methods in Ref.\cite{zin-borel}. In fact
this is acceptable because near the critical point the theory is highly non-perturbative and thus higher orders like $_{3}F_{2}$, $_{4}F_{3}$ and $_{5}F_{4}$ are expected to give more accurate result for the critical coupling.

\begin{table}[pth]
\caption{{\protect\scriptsize {The Hypergeometric resummation $_{2}F_{1}$,
$_{3}F_{2}$, $_{4}F_{3}$ and $_{5}F_{4}$ for the ground state function of the
Hamiltonian in Eq.(\ref{Yang}) compared to the $150^{th}$ order of resummation
methods in Ref.\cite{zin-borel} and the $20^{th}$ order of the perturbation
series ($E_{per}$) from Ref.\cite{zin-borel}. Our resummation formulae all
show up $\mathcal{PT}$-symmetry Breaking and precision is improved using
higher orders. Our third ($_{3}F_{2}$), fourth ($_{4}F_{3}$) and fifth
($_{5}F_{4}$) orders are showing results with competitive precision to the
$150^{th}$ order of resummation methods in Ref.\cite{zin-borel}.}}}%
\label{tab:wxiresum}
\scalebox{0.7}{
\begin{tabular}{|l|l|l|l|l|l|l|}
\hline
$J$                       & $_{2}F_{1}$                  & $_{3}F_{2}$                  & $_{4}F_{3}$         & $_{5}F_{4}$ & $E^{150}_{0}$ \cite{zin-borel}         & $ E_{per}$     \\ \hline
$-2^\frac{4}{5}$ & 0.289111-0.345157 i& 0.388417-0.328504i& 0.394688-0.3560448i & 0.388902-0.358021i  & 0.389 8-0.3644i    & 2.52955           \\ \hline
-1                   & 0.229176 - 0.029543 i          & 0.19719967          & 0.19580355    & 0.195741   & 0.1957508&            0.19574 \\ \hline
$-5^\frac{-4}{5}$                  & 0.282836           & 0.282700           & 0.282699& 0.28269926   & 0.282699&             0.282699 \\ \hline
$-21.6^\frac{-4}{5}$                  & 0.342161           & 0.342158         & 0.342158& 0.342158   & 0.342158 &             0.342158 \\ \hline
\end{tabular}}\end{table}

To conclude, we stressed the recently introduced Hypergeometric Resummation
algorithm \cite{Prl}. We realized that when applying the algorithm to the
$\mathcal{PT}$-symmetric $i\phi^{3}$ field theory it gives accurate results
for a range of the coupling values but for very large coupling values the
results are deviated from expected ones either from exact calculations or from
strong coupling limits where both are known from the literature. We expected
that the reason behind this is that the four parameters of the Hypergeometric
function are all predicted from the first four perturbative terms in the
divergent series of the ground state energy and thus the resummed function has
no guidance for strong coupling values. In fact, there exist well known
techniques to obtain the strong coupling expansion of a physical quantity
either for the quantum mechanical case and some times for quantum field
cases \cite{zin-borel, bend-strong, keinert-strong}. Accordingly, we suggested
to feed the Hypergeometric function with two parameters that can be predicted
from the strong coupling behavior and the other two parameters from the first
two terms in the perturbation series. In that way we obtained a Hypergeometric
function that bears information from weak coupling as well as strong coupling
behaviors and thus expected to give accurate results for the whole range of
the coupling space.

 We tested the algorithm by obtaining a very precise value for the critical exponent $\nu$ of the $O(4)$ field theoretic model 
from the first order in perturbation series and the asymptotic strong coupling data as input. Up to the best of our knowledge, this is the first time to get such accurate result from that low order of perturbation series.   This result assures  that the effect of involving exact parameters from strong coupling expansion shall accelerate the convergence of the resummation function toward exact results.

 We showed that the extension of the algorithm is direct and one can use as a resummation function the generalized
Hypergeometric function $_{p}F_{q}(a_{1,}a_{2},....a_{p};b_{1},b_{2,}.......b_{q};-\sigma z)$ where the parameters $a_{i}$ are all obtained exactly
from the strong coupling expansion of the physical quantity under consideration. In fact, this algorithm reduces the non-linearity issue which one faces when trying to find all the parameters from just perturbation series as it lowers
the number of equations in the parameters by a factor of half. Besides, one
can get an equivalent set of equations where are all linear in the parameter
$\sigma$ as well as having only the power one in each parameter. Also, the
algorithm guarantees reliable results even for very large coupling values.

We tested the idea for the anharmonic oscillator for $g=50$ and found that our second order ($\,_{2}F_{1}$) result
is better than the $24^{th}$ order for Borel- Pad$\acute{e}$ ($BP_{12-12}$) but is not
of same precision as the $25^{th}$ order of the generalized Hypergeometric (Meijer-G approximants)
algorithm in Ref.\cite{Prd-GF}. However, our fourth order calculation
$\,_{4}F_{3}(a_{1,}a_{2},....a_{4};b_{1},b_{2,}.......b_{3};-\sigma z)$ shows
good results in comparison with exact ones.

Since our main problem is to resum the ground state function of the
$\mathcal{PT}$-symmetric $i\phi^{3}$, we tested the prediction of the modified
algorithm and found that the second order calculation $\, _{2}F_{1}%
(a_{1,}a_{2} ;b_{1};-\sigma z)$ gives accurate results compared to exact
results for a wide range of the coupling. Of course as the coupling increases
our predictions go better than those from the original version of the
algorithm as explained above. The modified algorithm introduced here is
successful to reproduce the well known limit of the ground state energy as
$g\longrightarrow\pm\infty$. Also, the precision is improved when we use
higher order where the fourth order predictions give competitive results
compared to the $150^{th}$ order of the resummation algorithms used in
Ref.\cite{zin-borel} (Table \ref{tab:wgresum}).

The $\mathcal{PT}$-symmetry breaking of the Yang-Lee model has been tested
where the perturbation series can not account for it at any order. The second
order shows good agreement with $20^{th}$ order of the perturbation series of
the ground state energy but far from the critical region as well as large
values of the coupling, the perturbation series fails to give reliable
results. While the second order of our calculation gives reasonable results
and accounts for $\mathcal{PT}$-symmetry breaking at a negative coupling as it
was predicted  by the work in Ref.\cite{zin-borel}, it predicts smaller (in value)
critical coupling. The accuracy is highly increased in using higher orders
where the third, fourth and fith orders showed great results compared to the
to the $150^{th}$ resummation algorithm used in Ref.\cite{zin-borel} (Table
\ref{tab:wxiresum}).

The algorithm introduced here gives accurate results with less effort as one
is not in a need to obtain more than the second order of the strong coupling
expansion of a physical quantity and can then conclude all the strong coupling
parameters in the series since the coefficients in the strong coupling
expansion do not matter here. We think that this might be the most simple,
accurate and time saving resummation algorithm as it uses the already summed
huge set of functions $_{p}F_{q}(a_{1,}a_{2},....a_{p};b_{1},b_{2,}%
.......b_{q};-\sigma z)$ where the variety of parameters can fit with huge
number of problems in physics.

\end{document}